# Poisson's CDF applied to Flexible Skylines

Jaime Pons Garrido

Politecnico di Milano
Milan, Italy
jaime.pons@mail.polimi.it

**Abstract**

*The evolution of skyline and ranking queries has created new archetypes like flexible skylines, which have proven to be an efficient method to select relevant data from large datasets using multi objective optimization. This paper aims to study the possible applications of Poisson distribution mass function as a monotonic scoring function in flexible skyline processes, especially those featuring schemas whose attributes can be translated to constant mean rates. Moreover, a method to express user's requirement by means of the F-dominant set of tuples will be proposed using parametrical variations in F [1], simultaneously, algorithm construction and potential applications will be studied.*

## 1 Introduction

Dealing with data, especially when the amount is big, is a complex matter at several levels. In this regard, selecting significant relevant data from a large dataset is one of the processes that have hoarded many investigation resources to create a method to perform such operations in DBMS. Ranking and skyline queries have been de facto solutions for addressing such problems for long periods of time.

Ranking queries (or top-k queries) propose a weighted scoring system in which the query's output is the result of a ranking in which each data tuple is ranked according to a weighted scoring function and ordered in consequence [11]. The advantages brough by this model are many, first it allows one to control the cardinality of the result desired (top-k) by selecting the k-tuples with best score. Moreover, the weighted function allows users to express their criteria or preferences by modifying weights.

On the other hand, this type of query has major disadvantages in terms of efficiency and scalability since a weight change means all the previous computations are invalid. This may lead to several repetitions of the whole process until appropriate results are provided, thus, potentially wasting system resources in rankings that eventually will be obsolete.

Skyline queries implement a different approach, instead of grouping parameters and compare scalar values, tuples are treated like vectors and a dominance criterion is defined [2]. This criterion stablishes that a tuple is dominant over another if it contains at least one attribute greater than the same exact attribute of the other tuple and all the other ones are greater or equals than their counterparts. To be precise, let $t1$, $t2$ be tuples:

$$\text{SKY(R)} = \{t1 | \forall t2 \in R, \exists a_i\, t1(a_i) > t2(a_i) \land \forall a_i, i = 1, \dots n, t1(a_i) \geq t2(a_i)\} \quad (1)$$

This solution provides much better scalability since all computations are susceptible to be reused plus it does not maintain an order, instead it creates two subsets of dominated and non-dominated tuples. In return, its weaknesses match ranking strengths by not having control over the cardinal of the non-dominated set of tuples, very limited expression capabilities regarding user preferences, plus $O(n^2)$ efficiency.

## 1.1 Flexible skyline queries

Flexible skyline, also known as F-Skyline, queries intend to provide an in between solution able to be computationally affordable but also provide certain control over cardinality and user preferences. If Skyline queries were based on Pareto's Dominance concept [3,12], this new approach intends to combine it with the method applied in top-k queries.

Let r be a relational schema belonging to DBMS and let $A = (A_1, A_2, \dots, A_n)$ be the set of normalized ([0,1] - valued) attributes related to r. Let $\mathbb{M}$ be the family of all monotonic scoring functions ($f: [0,1] \to \mathbb{R}$), like the one that may be used in ranking queries. Flexible skylines now define the concept of $\mathbb{F}$ -dominance [1] as a Pareto Dominance applied to $e_i, e_j \in r$ where $e_i$ will dominate $e_j$ if and only if $e_i$ attributes have always better or equal scores than $e_j$ attributes with respect to the function subset $\mathbb{F} \subset \mathbb{M}$.

This approach allows computational efficiency since modifying the subset $\mathbb{F}$ only requires computing the additional functions added and deprecate those removed, but the ones that remain after the modification do not require additional computations. Moreover, user priorities or preferences may be expressed through constraint inclusions in the set $\mathbb{F}$ and cardinality of the result is able to be reduced by considering specific output subsets such as ND and PO [1,2,4].

DEFINITION 1.1: Let $\mathbb{F}$ be a set of monotone scoring functions. A tuple t $\mathbb{F}$-dominates another tuple s ≠ t, denoted by $t \prec_\mathbb{F} s$, if and only if $\forall f \in \mathbb{F}, f(t) \le f(s)$.

DEFINITION 1.2: (ND). Let $\mathbb{F} \subseteq \mathbb{M}$ be a set of monotone scoring functions. The non-dominated flexible skyline of r with respect to $\mathbb{F}$, denoted by ND (r; $\mathbb{F}$), is defined as the following set of tuples:

$$\text{ND}(r; \mathbb{F}) = \{t \in r | \nexists s \in r. s \prec_\mathbb{F} t\} \quad (2)$$

DEFINITION 1.3: (PO). Let $\mathbb{F} \subseteq \mathbb{M}$ be a set of monotone scoring functions. The potentially optimal flexible skyline of r with respect to $\mathbb{F}$, denoted by PO (r; $\mathbb{F}$), is defined as:

$$\text{PO}(r; \mathbb{F}) = \{t \in r | \exists \mathbb{F} \in \mathbb{M}, \forall s \in r, s \ne t \to f(t) < f(s)\} \quad (3)$$

## 1.2 Poisson distribution

The Poisson Distribution is a discrete probability distribution that has been prominent in the statistics field due to its convenient capability to express the probability of occurrence for a determined number of events in a fixed interval, usually of time. Poisson requires events to occur at constant mean rate and independently of the last event time of occurrence.[5]

DEFINITION 1.4: *(Mass function) Let X be a discrete random variable. X has Poisson distribution with $\lambda$ parameter, $X \sim Po(\lambda)$ if the probability of having k occurrences is, where $K \in \mathbb{Z}^+$:*

$$P(X = K) = \frac{\lambda^k e^{-\lambda}}{K!} \quad (4)$$

DEFINITION 1.5: (Cumulative Distribution Function) Let X be a discrete random variable. X has Poisson distribution with $\lambda$ parameter, $X \sim Po(\lambda)$ if the probability of having K or less occurrences is:

$$P(X \le K) = \frac{\Gamma(\lfloor K+1 \rfloor, \lambda)}{\lfloor K \rfloor!} \quad (5)$$

Where $\Gamma$ represents the incomplete gamma distribution.

After this brief introduction about the Poisson distribution one can realize that Poisson and, in general, all probabilistic distributions have a parametrical family in their CDF. In Poisson's case this family depends on $\lambda$ as a unique parameter. We provide a previous definition from the F-skyline theoretical development [1,2].

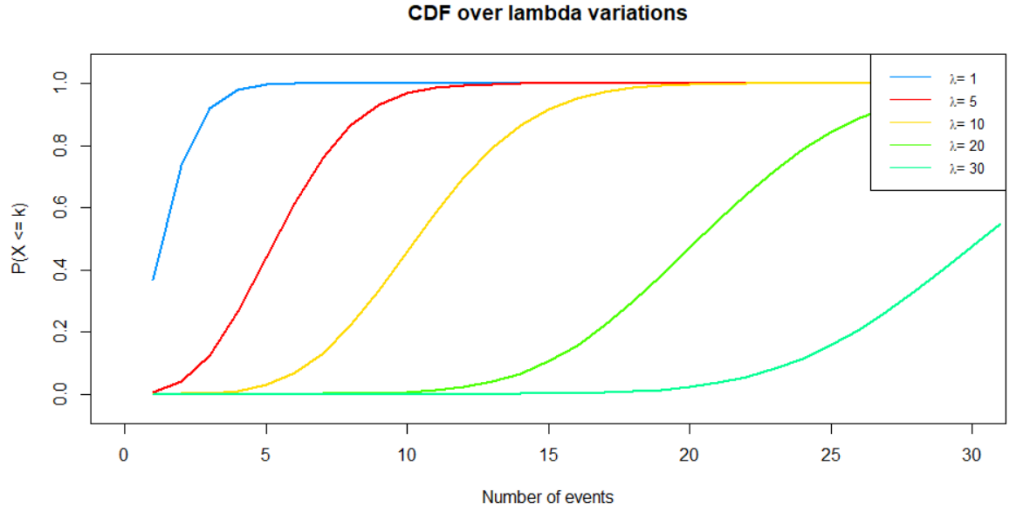

Figure 1: Poisson CDF iterating on several lambda values

DEFINITION 1.6: A set F of scoring functions is said to be tuple-distinguishing if the following holds:

$$\forall t, s \in [0,1]^d, t \neq s \rightarrow \left(\exists f \in \mathcal{F}. f(t) \neq f(s)\right) \qquad (6)$$

PROPOSITION 1.1: Let $P_{(\alpha 1 \ldots \alpha n)}$ be the subset conformed of Poisson CDF functions depending on an arbitrary n number of parameters $\rightarrow P_{(\alpha 1 \ldots \alpha n)}$ is also a subset of monotonic (non-decreasing) functions where $f \in P_{(\alpha 1 \ldots \alpha n)}$ is of the form $f: \mathbb{R} \rightarrow [0,1]$ plus it is tuple-distinguishing.

*Proof:*

*To prove that P is a subset of monotonic, more precisely, non-decreasing functions. It suffices with reasoning that the derivative of a Poisson CDF is its mass function. Following Def 1.4: $P(X = k) = \frac{\lambda^k e^{-\lambda}}{k!}$. It is trivial to check it is a coefficient between two positive factors that cannot be null. Therefore, CDF derivative is always positive which implies P is a subset of non-decreasing functions.*

*The fact that CDF is non-decreasing implies $\forall t, s, t \neq s \rightarrow f(t) \neq f(s)$, where f is a CDF. If we consider t and s such as $t \neq s$ composed by two suitable attributes but include the condition $f(t) = f(s)$ each we have the following:*

$$P(X \leq s[1]) + P(X \leq s[2]) = P(X \leq t[1]) + P(X \leq t[2]) \rightarrow$$
$$P(X \leq s[1]) - P(X \leq t[1]) = P(X \leq t[2]) - P(X \leq s[2])$$

*However, this implies the area integrated between s[1] and t[1] which is the increased mass of probabilistic density between both points is equal to the one between t[2] and s[2]. And this must fulfill for all possible Poisson distributions included in the family.*

*Considering that Poisson CDF derivative (Mass function) is always positive, but certainly not constantly positive. It can be stated that increases along the different areas of the CDF are not constant which means that the areas enclosed*

*between two different pairs of point, may be equal for a particular case but that equality cannot be extended if parametrical values change since derivatives will do too and not at the same rate. Implying:*

$$P(X \leq s[1]) - P(X \leq t[1]) \neq P(X \leq t[2]) - P(X \leq s[2])$$

*An equivalent assumption could be derived considering K as fixed value and instead using t and s parameters as mean constant rates ($\lambda$).* ∎

Considering what has been previously stated, one could make the argument that CDFs are, indeed, viable parametrical candidates to be applied to F-skylines. This paper aims to provide an initial overview about the possible use of Poisson CDF in F-skylines processes, this englobes:

- A brief theoretical introduction to Poisson Distribution, its asymptotic behavior.
- An approach to the possible methods to use Poisson CDF in the already developed F-skylines.
- A discussion about its possible potential applications, followed by final conclusions regarding overall advantages and disadvantages of the implementation.

## 2 Poisson CDF applied to F-skyline queries

The application of Poisson distribution to F-skylines must be strongly linked to the need of modelling Poisson-like processes within those queries. Nowadays, selection processes, ranking computations and many other procedures where the use of F-skylines is a viable option depend more on dynamic to be anticipated, not considered once they become a fact, for example the number of orders that will arrive to a warehouse at a determined time. A simple adaptation to this phenomenon may consist in making different use of the attributes stored at a relational schema and the usual constraints introduced to adapt to the user's preferences to create scoring functions able to take into account this kind of factor.

The solution proposed follows the following structure: let r be a relational schema and $a_i \in r$ is an attributed belonging to the model. Now let $a_i$ be the mean (expected value) of a dynamical process relative to the schema. (i.e., Let r be modelling a restaurant, then $a_i$ may be the expected number of guests from 8pm to 9pm), obviously the value of the attribute has been obtained by statistical inference.

PROPOSITION 2.1: Let X be a Poisson random variable, then its expected value is $\lambda$
*Proof: Stated in reference [5] page 1.*

By proposition 2.1 we can see that if we were to model the random variable relative to $a_i$ as a Poisson distribution, the parameter lambda corresponding to it would be $a_i$. Therefore, the relationship between the attribute and the distribution already generates a CDF function parametric family as a function of a K parameter, obtaining a subset:

$$\mathcal{F} = \{P(X \leq K) | K \in [0, +\infty), X \sim Po(a_i)\} \qquad (7)$$

K parameter is intended as a user preferences-oriented parameter which expresses the number of occurrences for the treated phenomenon desired by the user.

## 2.1 Poisson CDR asymptotic behavior as a scoring function

PROPOSITION 2.2 Let be X a Poisson distributed variable with fixed $\lambda$, then $\lim_{K\to 0}(P(X \leq K) - P(X \leq K+1)) = 0$ and $\lim_{K\to +\infty}(P(X \leq K) - P(X \leq K+1)) = 0$. *Proof: Stated in reference [6]*

The previous result indicates that the probability mass of a Poisson distribution is highly concentrated around $\lambda$, this means that when used as a scoring function, the expected behavior will be to experience significant scoring fluctuations between values close to the mean. However, for those values far enough, whether the number of occurrences is extremely high or low, the difference in scoring between then will not be relevant. [5, 6] A colloquial approach to this would be to expect that the phenomenon measured by the CDF will lose relevance in comparison to other attributes when located far away from the mean.

Regarding F-skylines, the usual implementation considers that tuples with lower scores are the best ones[1,2], however this is just a design choice and, obviously, depending on the results we expect there are several naïve transformations that can be applied to the scoring function to encourage different types of behaviors. For example, if higher scores are wished for a low constant K: $f := 1 - P(X \leq K) \equiv P(X > K)$ is a viable scoring function. On another case, if high scores are wished for K values close to the average number of occurrences expected ($\lambda$), function (8) can be used.:

$$f(X) = 1_{k<\lambda} \cdot P(X \leq K) + 1_{k\geq\lambda} \cdot P(X > K) \quad \text{where} \quad (8)$$

$$1_{k<\lambda} := \begin{cases} 1 \text{ if } k < \lambda \\ 0 \text{ else} \end{cases}$$

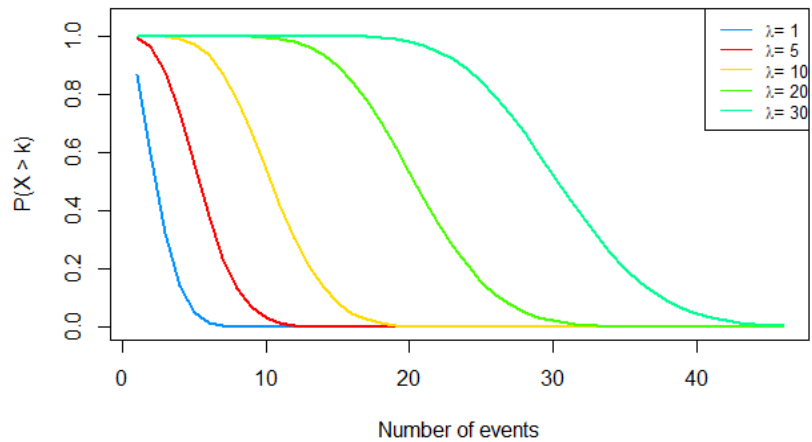

Figure 2: 1-CDF Poisson iterating on several lambda values

This function formulated like this may be understood as a non-monotonic function, however it is the addition of 2 monotonic functions that are independent and do not overlap, thus it can be decomposed therefore not supposing a problem for the evaluation of the score.

There is one last relevant problem to be addressed regarding CDFs application to F-skylines. If $\mathbb{M}$ is stated as the family of monotonic scoring functions ($f:[0,1] \to \mathbb{R}$) and $\mathbb{F} \subset \mathbb{M}$. We can directly see that Poisson CDF has inversed domains.

## 2.2 Domain inversion between Poisson CDF and ꟻ

There are two possibilities to address the domain inversion between both sets of functions. The first one implies limiting the CDF domain using the asymptotic behavior previously explained and the other one is simple, use the inverse of the CDF function ($CDF^{-1}$).

- The first solution tackles both domain and codomain problems, being the first one the most problematic. Domain of a CDF states that K that can be any real positive number, however the results in Proposition 2.1 indicate that for high K values the increment in CDF dismisses tending to 0. Thus, if chosen a C (constant) high enough, we can turn our domain $K \in \mathbb{R}^+$ into $K \in [0, C]$ which is limited. Afterwards, there is no possibility to reduce it to [0,1] without heavily altering the function's behavior.

  Now regarding the codomain problematic, having a codomain of [0,1] is not such a big issue since actually many scoring functions act in this way. However, if there are functions in ꟻ that score values over 1 and there is a proportional decompensation, multiplying the results of the CDF times a scalar factor suffices to bypass the problem. It could be understood as applying a weight to the function.

- Calculating $CDF^{-1}$ directly solves all domain problems, but changes the meaning of the distribution, which now does not calculate an accumulated probability but given one it returns its associated value K. This kind of function may also have several potential use cases.

These modifications are important to take into consideration, moving onwards Poisson CDF will be kept as the function to study without applying these changes for simplicity issues.

## 2.3 Computation procedures

Let the following result be considered from [1,2]:

PROPOSITION 2.3: For any set ꟻ of monotone scoring functions, $PO(r, ꟻ) \subseteq ND(r, ꟻ) \subseteq SKY(r)$
*Proof: Stated in reference [1], section 3.1, Prop 1.*

Two algorithms proposed by Ciaccia & Martinenghi [1] (`SVE1F` and `POND`) which can compute ND and PO subsets respectively, have been considered with the purpose of calculating those subsets using Poisson CDF. This only implies including an optimization feature able to take advantage of Poisson asymptotic behavior. The following subset type ꟻ will be considered:

$$ꟻ = \{P(X > w1) + a2 \cdot w2 + \cdots + an \cdot wn | X \sim Po(a1), wi \in [0,1] \forall i \neq 1, w1 \in [0, +\infty]\} \quad (9)$$

PROPOSITION 2.4: Let X be a random variable of the form $X \sim Po(\lambda)$. This implies Var $(X) = \lambda$
*Proof: Stated in reference [5], page 5.*

The modification in `SVE1F` takes place when the algorithm starts calculating the scores of a particular tuple in the ꟻ subset and it needs to calculate a score based with Poisson CDF as scoring function. Moreover, consists of assigning asymptotic limits to K values when they fall of the range of two typical deviations from the mean. The mean in this case is $a1$ (9) which will be the constant mean rate that was stored in the schema attributes. The asymptotic limit introduced will vary if the asymptotic limit tends towards 0 or towards 1. The following lines provide pseudocode for the modification.

| |
|---|
| : compute $v = \sum_{i=2}^{d} w_i^{(l)} s[A_i]^p$ //more references about this operation in [1] |
| : **if** $w1 \in [a1 - 2\sqrt{a1}, a1 + 2\sqrt{a1}]$ **then** compute $v = P(X > w1) + v, X \sim Po(a1)$//Poisson CDF case |
| : **else if** $w1 \geq a1 - 2\sqrt{a1}$ **then** compute $v = 1 + v$//opposite case would appear if $P(X \leq w1)$ |

In the previous pseudocode, the first line calculates the sum of the non-Poisson parameters scores, the second one checks that w1 falls within $[a1 - 2\sqrt{a1}, a1 + 2\sqrt{a1}]$ range, further explained in the following paragraph, if it does, it calculates its value following Poisson CDF. Otherwise, when the condition is not met, there are two possibilities, if the value is higher than the range, the third line computes v = 1 directly. If it turns out the value is lower than the range the associated probability will tend towards zero, thus nothing is added.

The reason behind this implementation choice is that in most distributions a large percentage of the probability mass falls within the range of two typical deviations. This is a well-known fact further explained in references [7,8] and empirically supported by the 68-95-99.7 rule. By doing this we pretend to save computational power by not effectuating Poisson CDF calculations that will return values extremely close to zero or one. As previously stated, mass function is the derivative of the CDF so, if the values of the first are close to zero outside of the range of two typical deviations, we can assume that the increase in CDF value will be minimal, which makes feasible to substitute its value for an asymptotic limit.

To select the potentially optimal subset (PO), the process proposed in reference [1], section 5.2 is followed. In this process, ND(r, F) subset tuples discard any combination F-dominated by a convex combination of ND tuples as stated in Theorem 11(Potential optimality test) [1], section 4.2. The question to make now is if CDF requires a different criterion rather than the exposed in that theorem to proof a tuple is PO. Considering F still is a sum of weighted monotonic scoring functions with single attributes, Theorem 11) [1] should also be applicable to this case. POND algorithm is explicitly described in reference [1], section 5.2.

## 3 Potential applications

The inclusion of CDFs and more precisely Poisson's into F-skylines could be a potential improvement to adapt this process to different fields in which a stochastic processing of data is encouraged, therefore augmenting the probabilities of creating relational schemas whose attributes are suitable to be valued by a CDF. Some of these applications will be reviewed now and are also disclosed in [9, 10, 12]:

**Recommendation-based systems:** Those based on the concept of human-choice simulation are prone to incorporate this new type of F-subsets if they use the F-skyline procedure. This enables them to improve their response to user or environment interactions by consider statistically inferred measurements in conjunction to CDFs to evaluate the best response to a stimulus. For example, using Poisson distribution to retrieve the user's most performed interaction after K occurrences of an event calculating the probability of those K occurrences taking place and offering adjusted recommendations in consequence.

**Managerial systems:** Relying heavily on their capability to improve the efficiency of resource allocation within large companies or systems, if they use large database systems to select the best options to allocate or free resources, then applying CDF to their selection processes using F-skylines may result on a drastic improvement. These types of uses are wide ranged, for example:

- Public health system tracking the number of emergency ambulances arriving to each hospital during certain critical periods of time, may use this to compute efficient destinations for personal, or medical equipment by selecting ND elements of F-skylines.
- Security forces can compute at large scale optimal routes to patrol the streets with higher probability of a certain number of violent events happening during nighttime periods.

**Machine learning applied to ranking:** In this kind of system filtering relevant information is vital to achieve quality results. Many big data sets come within a lot of irrelevant information; therefore, using F-skylines as selection process to identify relevant data can help to speed up processes. In this context, being able to select tuples who have higher probability of a determined future behavior may be useful information to feed the system.

**User Oriented comparation systems:** Poisson distribution is especially useful measuring probabilities related with mass transit through a determined place. This supposes at differential factor when a system must reduce the viable alternatives available to perform a task or travel to a location.

Medical healthcare apps provide the option to schedule medical appointments in several locations. A dynamic phenomenon increasing its relevancy regarding these selections is crowd avoiding. By registering the most booked hours at every center, the app should be able to use Poisson CDF and constant mean rates to concede more priority to those centers without much transit probability at the time of the appointment. By reflecting that positive attribute as a stochastic probability in the scoring function.

The opposite type of use is also a possibility, for example when searching viable leisure activities which significantly improve as the number of concurrent users performing them increases.

*Example 3.1: In this example a medical application helps a client to select the best hospital to schedule an appointment considering distance and the number of concurrent patients visiting the hospital as factors to make a choice.*

*Let a simple relational schema r with two attributes be considered. $A_r$=(AverageFrequency, Distance). Where AverageFrequency is the average number of patients arriving to a hospital on a determined time interval and Distance is a [0,1]- valued parameter representing distance to the location. The dominance relation between elements of r is defined as in Definition 1.1 and subset F as:*

$$F = \{P(X > k_1) \cdot w_1 + Distance \cdot w_2 | X \sim Po(AveragePatients), w_i \in [0,1], \sum w_i = 1, w_i < w_{i+1}, k_i \in [0, \infty] \} \quad (10)$$

*A first parametrical choice is made when selecting $k_1$, in our case, $k_1$ = 8. Now, our main interest is obtaining ND and PO sets, a process detailed in [1], sections 4, 5. This must be done variating the possible constraints in set $w_i$. In this case, we have considered a small subset conformed by five hospitals.*

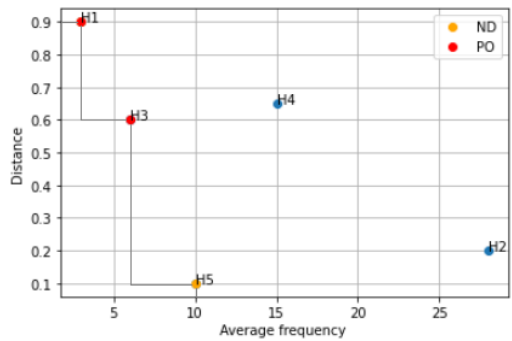

Figure 3: F-skyline regarding the most convenient medical centers nearby.

We can see that ND ≡{H1, H3, H5} and PO ≡{H1, H3}, Therefore, now the system is able to provide a better selection based on these results. This example also identifies some of the limitations of Poisson CDF regarding computation of ND and PO subsets. In the previous analysis the set $k_i$ was fixed and $w_i$ was the set of constraints subject to variation to find the subsets. The opposite could also be made, obtaining different results. However, the analysis gets much more complicated if both sets of constraints are subject to variation, that case is out of the scope of this paper and is practical utility may yet to be discussed.

## 4 Conclusions

The inclusion of new subsets of functions to already well formularized methodologies like flexible skylines may suppose an advantage in terms of expanding the capabilities of those. CDFs are natural candidates due to their suitability to form F subsets. Moreover, the inclusion of dynamic factors and statistical parameters into data structures makes more feasible to find indicators (i.e., mean, variance, etc.) in databases subject to F-skylines operations, therefore, being able to use CDF poses an intrinsic advantage in many more cases because it allows to consider probabilities associated to these indicators when the adequate distribution to model the events is selected.

Focusing on Poisson distribution it seems that including it into an $\mathcal{F}$ subset mixed with low-grade polynomial functions commonly used in F-skyline procedures [1,2] allows to keep using the standard algorithmic procedures. Plus, as discussed on Section 2, knowledge about asymptotic behavior and derivatives may be useful in optimization terms when the inserted values are extreme. In addition, Poisson unique characteristics, especially the one formulated in Proposition 2.1, provide unparalleled simplicity to store the relevant statistical indicators and a clear path to involve them as parametrical values once the Poisson CDF starts being treated like a scoring function within $\mathcal{F}$.

On the other hand, if other statistical distributions were evaluated for suitability, it would not be such a simple implementation due to the large variability between probabilistic distribution's properties. This would imply more expensive computations to find a valid parameter for those distributions while in Poisson's case the mean is acceptable. Moreover, there are still many unresolved matters to study regarding Poisson CDF, to mention some examples:

- ND and PO behavior in F subsets composed purely of Poisson CDFs and those which are mixed. How does the election of $k_i$ parameters affect the cardinality of both sets? Is it efficient to search ND and PO taking $k_i$ parameters into account?

- Perform empirical simulations of F-skyline processes including Poisson CDFs to evaluate their efficiency and complexity, followed by a comparation against other relevant methodologies able to offer similar results in terms of non-dominated tuples.

- Further study the application of F-skylines to build complex stochastic models like Markov Chains based on Poisson processes.

- Evaluate and compare Poisson's characteristics to identify other suitable distributions to be included into $\mathcal{F}$ subsets in meaningful ways regarding their capabilities as stochastic quantifiers.